\documentclass[reviewcopy]{elsart}
\usepackage{amsmath}
\usepackage{graphicx}
\usepackage{natbib}

\begin{document}

\begin{frontmatter}

\title{\textbf{Energy efficiency of information transmission by electrically coupled neurons}}

\author[first]{Francisco J. Torrealdea}
\author[second]{Cecilia Sarasola}
\author[first]{Alicia d'Anjou}
\author[first]{Abdelmalik Moujahid \corauthref{upv}}
\ead{jibmomoa@gmail.com}
\author[third]{N. V\'{e}lez de Mendiz\'{a}bal}

\address[first]{Department of Computer Science, University of the Basque
Country UPV/EHU, 20018 San Sebastian, Spain}

\address[second]{Department of Physics of Materials, University of the
Basque Country, 20018 San Sebastian, Spain}

\address[third]{Neuroimmunology Laboratory, Center for Applied Medical
Research, University of Navarra, Pamplona, Spain}

\corauth[upv]{Department of Computer Science, University of the
Basque Country, 20018 San Sebastian, Spain}

\begin{abstract}

The generation of spikes by neurons is energetically a costly
process. This paper studies the consumption of energy and the
information entropy in the signalling activity of a model neuron
both when it is supposed isolated and when it is coupled to
another neuron by an electrical synapse. The neuron has been
modelled by a four dimensional Hindmarsh-Rose type kinetic model
for which an energy function has been deduced. For the isolated
neuron values of energy consumption and information entropy at
different signalling regimes have been computed. For two neurons
coupled by a gap junction we have analyzed the roles of the
membrane and synapse in the contribution of the energy that is
required for their organized signalling. Computational results are
provided for cases of identical and nonidentical neurons coupled
by unidirectional and bidirectional gap junctions. One relevant
result is that there are values of the coupling strength at which
the organized signalling of two neurons induced by the gap
junction takes place at relatively low values of energy
consumption and the ratio of mutual information to energy
consumption is relatively high. Therefore, communicating at these
coupling values could be energetically the most efficient option.

\end{abstract}


\end{frontmatter}

\section{Introduction}
The relation between cerebral energy consumption and neuronal
activity was first suggested by Roy and Sherrington \citep{Roy}. A
neuron transmits information by depolarizing and repolarizing its
membrane to generate action potentials, what requires energy obtained from ATP produced from
glucose in the mitochondria. The rate of ATP generation depends on
multiple factors and if any of them causes the replenishment of
ATP supplies not be enough to satisfy the demand from the neuron
its refractory period will increase and the information will be
altered. The generation of actions potentials, or spikes, is
metabolically costly with energy demands tightly coupled to
spiking frequency \citep{Lennie,Smith} what makes the metabolic energy
required to maintain neural activity in a global scale very high
\citep{Clarke, Attwell, Laughlin, Siekevitz}. In humans, for
instance, the brain has only $2\%$ of the body mass and consumes
$20\%$ of the human metabolic energy \citep{Clarke} being a large
fraction of this total energy expended in the generation of the
firing sequences of action potentials that neurons use to
represent and transmit information \citep{Attwell}. The demand of
energy to generate these sequences of action potentials is so high
that energy supply seems to be a real constraint for neural coding
\citep{Laughlin} and, it has been suggested that nature, searching
a compromise between energy consumption and representational
capacity, might have developed energy efficient codes, that is,
codes that maximize the ratio of representational capacity to
energy expended \citep{Levy96, Levy02, Schreiber}. For instance, in the early visual system energy efficient coding could be a real biologically-based justification for sparse coding in the cortex and retinal ganglion cells \citep{Vincent}.

This paper approaches the problem of whether biological computation
optimizes energy use in the way neurons communicate. The evaluation
of the energy efficiency of the transmission requires both,
calculation of the amount of information transmitted and calculation
of the energy cost of the transmission. Quantitative mathematical
models have proved to be an indispensable tool in pursuing the goal
of understanding neuron dynamics \citep{Herz} and the study of
models showing the irregular spike bursting characteristic of real
neurons \citep{Hodgkin,FitzHugh,Hindmarsh85, Rose} has received much
attention
\citep{Rulkov,DeMonte,Ivanchenko,Venaille,Abarbanel,Huerta,Rosenblum,Belykh,
Hayashi,Lago,Yoshioka,Hasegawa,Nagai}. As these models are kinetic
models with no energy considerations, it could be of interest to
further develop them in such a way that they could be used to study
the relation between the dynamical properties of a neuron and
its energy implications. In Refs. \citep{Torrealdea, Torrealdea07}
we deduced for a three dimensional Hindmarsh-Rose neuron
\citep{Hindmarsh85, Rose} a function of its variables and parameters
that can be associated to the neuron as a real electrical energy.
This energy function was used to evaluate the energy consumption
of the neuron during its signalling activity. Our procedure to find a physical energy compatible with the dynamics of a dynamical system described by differential equations can be used to associate energies to many of the generally used models of neurons. Simple models with polynomial derivatives such as FitzHugh-Nagumo or Hindmarsh-Rose models are particularly apposite to associate to them an electrical energy function. Most of these models have been inspired in the work of Hodkking and Huxley but they do not conserve the clear physical meaning of the original work making it difficult to associate to them a physical energy. It is precisely here where our method can help. In this paper we deduce an energy function for a four dimensional Hindmarsh-Rose model. It is remarkable that this four dimensional energy turns out to be a natural extension of the one we found for the three dimensional case, what enhances the confidence in our result.

Energy efficient information transmission from the point of view that inputs are optimally encoded into Boltzmann distributed output signals has been analyzed in \citep{Balasubramanian}. An open question is the determination of the energy cost of generating the spike trains that codify each of the different output symbols. Our approach could provide a way to determine the energy cost of the generation of these spike trains.

Most of the cells in the nervous system are interneurons, that is,
neurons that communicate only to other neurons and provide
connection between sensory and motor neurons. Signals are
transferred from one neuron to another through synaptic junctions
which can be chemical or electrical. The rate of information
transmitted between two neurons can be quantified calculating the
mutual information between the corresponding trains of spikes of the
presynaptic and postsynaptic neurons \citep{Rieke}. Electrical
synapses are considered to be frequent and, it is believed, that
they provide a flexible mechanism for modifying the behavior of an
oscillatory neural network \citep{Connors, Kepler}. Most of the
electrical synapses are formed by gap junctions between neurons of
the same type, homologous gap junctions. Heterologous gap junctions
are less frequent \citep{Galarreta}. In this paper we analyze model
neurons of the same type and we refer to two neurons as identical
when they share the same set of parameter values and as nonidentical
when they differ in the value of some parameter.

A neuron responds to income signals from other neurons with
changes in its firing regime that modify its information capacity as
well as its average energy consumption. A natural way to propagate
information through a channel of neurons could be via partial or
total synchronization of the postsynaptic neuron to the signalling
pattern it receives from its presynaptic neighbor. For instance,
electrical synapses between AII amacrine cells and ON-cone bipolar
cells are considered essential for the flow of visual signals in the
retina \citep{Kolb} and temporally precise synchronization between
them of subthreshold membrane potential fluctuations has been
demonstrated \citep{Veruki2002a}. The degree of synchronization reached by the neurons conditions their capacity to transmit information and the energy consumption of their signalling activity. As the degree of synchronization is highly dependent on the coupling some coupling conditions may be more favorable than others for an energetically efficient transmission of
signals. In this work we investigate how this efficiency depends on
the type of coupling, unidirectional and bidirectional, and on the
values of the coupling strength when both neurons are coupled
electrically. We also investigate the role of the electrical
junction in the provision of the energy that the neurons require to
maintain their synchronized regime.

In Sec. 2 we summarize a procedure to find an energy function that
quantifies the physical energy associated to the states of a
generic model neuron described by differential equations. This
function can be used to quantify the consumption of energy of the
neuron in its different possible signalling regimes. We also
discuss the balance of energy when two generic neurons are coupled
electrically and quantify the contribution of the synapse to the
total energy required for both neurons to maintain the
synchronized signalling activity. This discussion is
particularized to the case of a four dimensional Hindmarsh-Rose
model of thalamic neurons for which analytical expressions of
energy consumption and synapse contribution are given. In Sec. 3 we
present some considerations relative to the way we have computed
the information entropy and the mutual information of two
electrically coupled neurons. In Sec. 4 we present computational
results. Firstly, results are given for the information entropy
rate and energy consumption of the different patterns of spike
trains that are generated by an isolated neuron at different
values of the applied external current. Secondly, results are
given for two neurons coupled electrically. Four cases have been
studied. Identical and nonidentical neurons coupled with unidirectional coupling and with bidirectional symmetrical coupling. For each studied case results of mutual information rates, energy
consumption, ratios of mutual information to energy consumption,
and relative weight of the synapse contribution of energy  are
presented and discussed. Finally in Sec. 4 we give a brief summary
and present our conclusions.

\section{Energy considerations}
In this section we quantify the energy required by a model neuron to
maintain its signalling activity. We analyze the energy requirements
when the neuron acts as an isolated oscillator and also the energy
aspects linked to the synaptic junction when two neurons are
electrically coupled. In order to quantify theoretically the energy
consumption of a model neuron we require an analytical expression of
the energy of the neuron in its different possible states. In Ref.
\citep{Sarasola} we described how to associate to a chaotic system a
function of its dynamical variables that can be formally considered a real physical energy of the system. By \emph{real physical energy} we mean that if a set of kinetic equations is considered a good model for the dynamical behavior of, for instance, a thalamic
neuron, then we must consistently consider the energy associated to it a good model for the energy implications of that dynamical behavior. We have tested the procedure with many electrical and mechanical systems always obtaining the correct energy. In \citep{Sarasola} an example is given for an oscillatory electric circuit. In the following section we very
quickly summarize the procedure described in Ref. \citep{Sarasola} which can be used
to find an energy function for a model neuron.

\subsection{Energy function associated to a model neuron}
Let us consider an oscillatory autonomous dynamical system represented by
\begin{math}
\mathbf{\dot x} = \mathbf{f(x)},
\end{math}
where $\mathbf{x} \in \Re^{n}$ and $\mathbf{f}:\Re^{n} \rightarrow
\Re^{n}$ is a smooth function, as the mathematical model of a
generic neuron. The velocity vector field $\mathbf{f(x)}$ can be
expressed as sum of two vector fields $ \mathbf{f(x)} = \mathbf{f}_c
(\mathbf{x}) + \mathbf{f}_d (\mathbf{x})$, one of them, $
\mathbf{f}_c (\mathbf{x})$, conservative containing the full
rotation and the other, $ \mathbf{f}_d (\mathbf{x})$, dissipative
containing the divergence \citep{Kobe}. Taking the conservative
vector field, the equation
\begin{equation}
\nabla H ^T \mathbf{f}_c (\mathbf{x}) = 0,
\end{equation}
where $ \nabla H^T $ denotes the transpose gradient of function $H$,
defines a partial differential equation from which a function
$H(\mathbf{x})$ can be evaluated. This function $H(\mathbf{x})$ is a
generalized Hamiltonian for the conservative part $\mathbf{\dot x}=
\mathbf{f}_c(\mathbf{x})$ as long as it can be rewritten in the form
$\mathbf{\dot x}= J(\mathbf{x})\nabla H $ where $ J $ is a skew
symmetric matrix that satisfy Jacobi's closure condition
\citep{Olver,Morrison}. If that is the case, we consider
$H(\mathbf{x})$ as an energy associated to the original system
$\mathbf{\dot x} = \mathbf{f(x)}$.  This energy is dissipated,
passively or actively, due to the dissipative component of the
velocity vector field according to the equation,
\begin{equation}
\dot H = \nabla H^T \mathbf{f}_d (\mathbf{x}).
\end{equation}

In Ref. \citep{Torrealdea} we used this procedure to find an energy
function for the well-known three variable Hinmarsh-Rose thalamic
model of a neuron. In the last part of this section we apply
 the same procedure to find and energy function for the four dimensional version of the model that
 was introduced by Pinto et al. in Ref. \citep{Pinto}. This energy function is used to evaluate the energy consumption of the neuron in isolation and also when it is connected to other neurons through electrical synapses. It provides the basis for all the computational results presented in this work.

\subsection{Electrically coupled neurons. Energy contribution from the synapse}

\begin{figure}
\begin{center}
\includegraphics[width=0.60\textwidth]{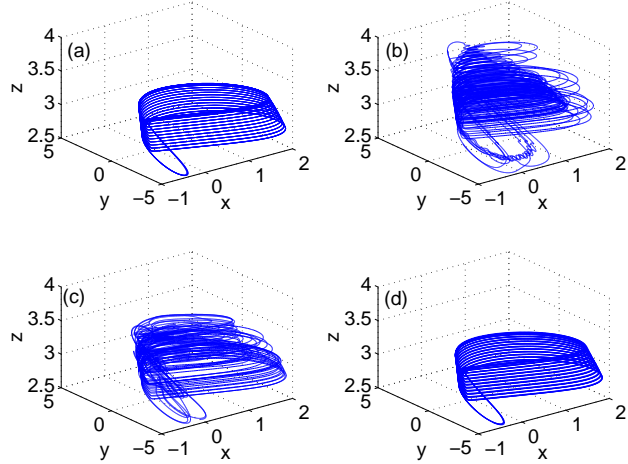}
\\
\end{center}
\caption{Projection on the $x,y,z$ axes of the attractor of a postsynaptic four dimensional Hindmarsh-Rose neuron forced by an electrical unidirectional synapse, Eqs. 15, to follow another identical neuron. Values of the coupling strength (a) $k=0$, (b) $k=0.25$, (c) $k=0.7$, (d) $k=1$. } \label{fig1}
\end{figure}

In this section we analyze the balance of energy required to maintain the signalling activity of two model neurons coupled by an electrical synapse. Let us consider the oscillating neurons $\mathbf{\dot {x}}_1 = \mathbf{f}_1(\mathbf{x}_1)$ and $\mathbf{\dot {x}}_2 = \mathbf{f}_2(\mathbf{x}_2)$ coupled electrically according to the scheme,

\begin{equation}
\begin{array}{ll}
\mathbf{\dot {x}}_1 = \mathbf{f}_1(\mathbf{x}_1)+\mathbf{K}_1(\mathbf{x}_2-\mathbf{x}_1) \\
\mathbf{\dot {x}}_2 = \mathbf{f}_2(\mathbf{x}_2)+\mathbf{K}_2(\mathbf{x}_1-\mathbf{x}_2), \\
\end{array}
\end{equation}
where $\mathbf{f}_1,\mathbf{f}_2: \Re ^n \to \Re ^n $ are smooth functions, $ \mathbf{K}_1,\mathbf{K}_2 \in \Re^n\times\Re^n $ are diagonal matrices representing the coupling strength with entries $k_1^i, \, k_2^i\geq 0$ \, $i=1,...n$, and  $ \mathbf{x}_1, \, \mathbf{x}_2 \in \Re ^n $ indicate the state of the coupled neurons. Note that the terms $ \mathbf{K}_1(\mathbf{x}_2-\mathbf{x}_1)$ and $\mathbf{K}_2(\mathbf{x}_1-\mathbf{x}_2)$ represent the gap junction that connects both neurons. These two terms, taken together, represent a potentially bidirectional electrical junction with selective ion channels of different conductances depending on the coupling matrices $\mathbf{K}_1$ and $\mathbf{K}_2$.

The signalling activity of a neuron consists of the generation of action potentials with different patterns of repetitive firing or bursting. This oscillatory behavior, when represented in the  phase space, makes the temporal evolution of the neuron remain confined to an attractive region which is characteristic of its dynamics. When two neurons are coupled their respective dynamics change although still remaining confined to attractive regions in the phase space, see Fig. 1 for the particular case analyzed in Section 2.3.2 of the paper. The nature of the coupled oscillatory regime of each neuron depends on the particular values of the coupling matrices $\mathbf{K}_1$ and $\mathbf{K}_2$. As the temporal trajectory $ \mathbf{x}_1(t) $ remains confined to an attractive region of the phase space, the long term net average energy variation along that trajectory of the system  $ \mathbf{\dot {x}}_1 = \mathbf{f}_1(\mathbf{x}_1)+\mathbf{K}_1(\mathbf{x}_2-\mathbf{x}_1)$, neuron and synapse, is zero. That is,
\begin{equation}
\langle\nabla H_{f_1}^T \, \mathbf{f}_1 (\mathbf{x}_1)\rangle +
\langle\nabla H_{f_1}^T \, \mathbf{K}_1(\mathbf{x}_2 - \mathbf{x}_1 )\rangle = 0 ,
\end{equation}
where the brackets represent averaging on the attractor and $ \nabla H_{f_1}^T
$ denotes the transpose gradient of the energy function of the neuron $ \mathbf{\dot{x}}_1=\mathbf{f}_1(\mathbf{x}_1) $. The same argument applies to system $ \mathbf{\dot {x}}_2 = \mathbf{f}_2(\mathbf{x}_2)+\mathbf{K}_2(\mathbf{x}_1-\mathbf{x}_2)$.

The first term of Eq. (4) can be associated with the variation of the energy of the first neuron through its membrane  and the second term with the variation of its energy through the synapse. Thus, for each of the coupled neurons, $  i=1,2$, the following balance of energy applies on average,
\begin{equation}
\langle\dot{H}\rangle_m^i+\langle\dot{H}\rangle_s^i=0,
\end{equation}
where $\langle\dot{H}\rangle_m^i$ and $\langle\dot{H}\rangle_s^i$ stands for the average energy variation of neuron $i$ through its membrane and synapse respectively.

According to Eq. (4), if $\mathbf{K}_1=0$, the average energy variation of the energy of neuron one through its membrane is zero, $\langle\dot{H}\rangle_m^1=0 $. The same applies to neuron two if $\mathbf{K}_2=0$. In other words, if a neuron does not receive signals from any other neuron the energy it obtains through the membrane, $ \langle\dot{H^+}\rangle_m^i, $ is perfectly balanced by its dissipation of energy through the membrane, $\langle\dot{H^-}\rangle_m^i $. Nevertheless, according to Eq. (5), if a neuron is signalling forced by signals arriving  from another neuron through a gap junction, it is the global average, membrane plus synapse, what is zero. The average variation of energy through the membrane is no longer zero. Therefore, when two neurons are coupled a contribution of energy from the synapse is required to maintain their cooperative behavior. This fact is a consequence of the forced oscillatory regime induced by the synapse and it is to be expected that the relative weight of the contribution of energy from the synapse to the energy balance of the coupled neuron be dependent on the strength of the synapse itself.

For each of the neurons, $i=1,2$, Eq. [5] can be rewritten as,
\begin{equation}
\langle\dot{H^+}\rangle_m^i=-\langle\dot{H^-}\rangle_m^i-\langle\dot{H}\rangle_s^i,
\end{equation}
where $\langle\dot{H^+}\rangle_m^i$ stands for the average of the positive part of the energy derivative, i.e., the energy income rate through the membrane, and $\langle\dot{H^-}\rangle_m^i$ for the average of the negative part of  the energy derivative, i.e., the energy dissipation rate through the membrane. Equation [6] emphasizes the fact that the total average income of energy through the membrane of the neuron equals its dissipation trough the membrane plus a net flow of energy in the synapse. From Eq. [6] the relative weight, $Sw$, of the contribution of the synapse to the total energy income would be,
\begin{equation}
Sw=\frac{\langle\dot{H}\rangle_s^i}{\langle\dot{H^+}\rangle_m^i}.\\
\end{equation}
This expression will be used later on in the paper to quantify the relative contribution of the synapse in the different coupling conditions studied in this work.\\

For the particular case of identical neurons and bidirectional gap
junctions $\mathbf{f}_1\equiv \mathbf{f}_2=\mathbf{f}$ and
$\mathbf{K}_1\equiv \mathbf{K}_2=\mathbf{K}$, the total average
energy variation of neurons one and two in the synapse is,
\begin{equation}
\langle\dot{H}\rangle_s=\langle\dot{H}\rangle_s^1+\langle\dot{H}\rangle_s^2,
\end{equation}
where
\begin{equation}
\langle\dot{H}\rangle_s^i=\langle\nabla H_f(\mathbf{x}_i)^T \, \mathbf{K}(\mathbf{x}_j - \mathbf{x}_i )\rangle,
\end{equation}
with $i,j=1,2\,;  \, \, i\neq j$. Equations (9) are symmetrical
with respect to an exchange of variables $\mathbf{x}_i$ and
$\mathbf{x}_j$ and as both neurons are identical and
undistinguishable $\langle\dot{H}\rangle_s^1$ and
$\langle\dot{H}\rangle_s^2$ must be equal and, therefore, the
total energy variation in the synapse is $2 \, \langle\nabla
H_{f}^T \, \mathbf{K}(\mathbf{x}_j - \mathbf{x}_i )\rangle $.  On
the other hand, as both neurons are identical and the gap junction
bidirectional and symmetrical, they are energetically identical
with relation to the synapse and, on average, there cannot be any
net flow of energy from one neuron to the other through the
synapse. As the net average energy variation at the synapse site
is not zero the gap junction itself must act as a source or sink
of energy for both neurons. Note that the degree of
synchronization reached, measured in terms of the norm of the
error vector $\mathbf{e}=\|\mathbf{x}_j-\mathbf{x}_i\|$ , will
condition the magnitude of this contribution of energy. If the two
neurons are identical, and the strength of the coupling large
enough, the synchronization error goes to zero and, therefore,
signalling transmission between identical neurons in complete
synchrony occurs with no energy contribution on average from the
synaptic junction.

\subsection{The four dimensional Hindmarsh-Rose neuron}

The Hindmarsh-Rose model of a thalamic neuron \citep{Hindmarsh85,
Rose} is a qualitative three dimensional model which is widely used
in the study of neuron dynamics because it can produce several modes
of spiking-bursting activity, including a regime of chaos, that
appear similar to those seen in biological neurons. The model,
although qualitative, is not unrealistic. Rose and
Hindmarsh in a series of papers \citep{Rose89a,Rose89b,Rose89c}
showed how a Hodking-Huxley like model,
based on ionic currents that can be related to experimental
recordings, is derived from it. However its parameter space for chaotic behavior is much
more restricted than what is observed in real neurons
\citep{Selverston,Pinto}. The chaotic behavior is greatly expanded
by incorporation of a fourth slow variable that increases the
realism of the description of slow Calcium currents. This four
dimensional model produces simulations of intracellular activity
which are even more similar to the biological observations
\citep{Selverston,Pinto}. In this paper we represent a single neuron
by the four dimensional extension of the original Hindmarsh-Rose
model which is described by the following equations of movement:
\begin{equation}
\begin{array}{ll}
\dot{x}= \mbox{a}y+\mbox{b}x^2-\mbox{c}x^3-\mbox{d}z+\xi I,\\
\dot{y}=\mbox{e}-\mbox{f}x^2-y-\mbox{g}w,\\
\dot{z}=\mbox{m}(-z+\mbox{s}(x+\mbox{h})),\\
\dot{w}=\mbox{n}(-\mbox{k}w+\mbox{r}(y+\mbox{l})).
\end{array}
\end{equation}

In the model variable $x$ is a voltage associated to the membrane
potential, variable $y$ although in principle associated to a
recovery current of fast ions has been transformed into a voltage,
and variable $z$ is a slow adaptation current associated to slow
ions. These three first equations constitute the standard three
dimensional model. Variable $w$ represents an even slower process
than variable $z$ and was introduced because a slow process such as
the calcium exchange between intracellular stores and the cytoplasm
was found to be required to fully reproduce the observed chaotic
oscillations of isolated neurons from the stomatogastric ganglion of
the California spiny lobster \emph{Panulirus interruptus}
\citep{Pinto}. Parameter $I$ is a external current input. The time
variable of the model is dimensionless. For the numerical results of
this work we fix the parameters to the values $\mbox{a}=1$,
$\mbox{b}=3.0 \, (\textrm{mV})^{-1}$,
$\mbox{c}=1\,(\textrm{mV})^{-2}$, $\mbox{d} =0.99\,
\textrm{M}\Omega$, $\xi=1\, \textrm{M}\Omega $, $\mbox{e}=1.01 \,
\textrm{mV}$, $\mbox{f}=5.0128 \, (\textrm{mV})^{-1} $,
$\mbox{g}=0.0278\, \textrm{M}\Omega $, $\mbox{m}=0.00215$,
$\mbox{s}=3.966\, \mu \textrm{S} $, $\mbox{h}=1.605 \, \textrm{mV}$,
$\mbox{n}=0.0009$, $\mbox{k}=0.9573$, $\mbox{r}=3.0 \, \mu
\textrm{S} $, $\mbox{l}=1.619 \, \textrm{mV}$. These numerical
values refer to $\textrm{cm}^2$ and are the same that have been used
in Ref. \citep{Pinto}. Both the three dimensional and four
dimensional models have regions of chaotic behavior, but the four
dimensional model has much larger regions in parameter space where
chaos occurs \citep{Pinto}.

\subsubsection{Energy consumption when signalling in isolation}

 In the Hindmarsh-Rose model given by Eq. (10) the
 vector field $\mathbf{f(x)}$ can be expressed as sum of the following vector fields,
\begin{equation}
 \mathbf{f}_c(\mathbf{x})  = \begin{pmatrix}
  \mbox{a}y-\mbox{d}z \\
  -\mbox{f}x^2-\mbox{g}w \\
   \mbox{ms}x \\
   \mbox{nr}y
\end{pmatrix} \quad   \text {and} \quad \mathbf{f}_d(\mathbf{x})  = \begin{pmatrix}
  \mbox{b}x^2-\mbox{c}x^3+\xi \mbox{I}  \\
  \mbox{e}-y  \\
  \mbox{msh}-\mbox{m}z\\
  \mbox{nrl}-\mbox{nk}w
\end{pmatrix}.
\end{equation}
As it can be observed $ \mathbf{f}_c(\mathbf{x}) $ is a divergence free vector that
accounts for the whole rotor of the vector field $ \mathbf{f}(\mathbf{x}) $, and $\mathbf{ f}_d(\mathbf{x}) $ is a gradient vector that carries its whole divergence. Consequently,
the energy function $H(x,y,z,w)$ will obey the following partial
differential equation,
\begin{equation}
(\mbox{a}y-\mbox{d}z)\frac{{\partial H}}{{\partial x}}-(\mbox{f}x^2+\mbox{g}w)\frac{{\partial
H}}{{\partial y}} + \mbox{ms}x\frac{{\partial H}}{{\partial z}}+ \mbox{nr}y\frac{{\partial H}}{{\partial w}} = 0,
\end{equation}
which has the cubic polynomial solution
\begin{eqnarray}
H(x,y,z,w)=\frac{p}{\mbox{a}} \left(\frac{2}{3}\, \mbox{f} x^3 +
\frac{\mbox{msd}-\mbox{gnr}}{\mbox{a}} x^2 + \mbox{a}y^2 \right) + \nonumber \\
\frac{p}{\mbox{a}} \left(\frac{\mbox{d}}{\mbox{ams}}(\mbox{msd}-\mbox{gnr})z^2-2\mbox{d}yz+2\mbox{g}xw \right)
\end{eqnarray}
where $p$ is a parameter. As in the model time is
dimensionless and every adding term in Eq. (13) has dimensions of
square voltage, function $H$ is dimensionally consistent with a
physical energy as long as parameter $p$ has dimensions of
conductance. In this paper we fix parameter $p$ to the arbitrary
value $p=-1\, \textrm{S}$. The minus sign has been chosen to make consistent the outcome of the model with the usual assumption of a demand of energy associated with the repolarization period of the membrane potential and also with its refractory period (see Fig. 2).

Note that if parameter $g$ is set to zero the four dimensional system given by Eqs. (10) reduces itself to the standard three dimensional model, as variable $w$ becomes uncoupled, and Eq. (13) reduces to,
\begin{displaymath}
H(x,y,z) =\frac{p}{\mbox{a}} \left(\frac{2}{3}\, \mbox{f} x^3 + \frac{\mbox{msd}}{\mbox{a}} x^2 + \mbox{a}y^2+\frac{\mbox{d}^2}{\mbox{a}}z^2-2 \, \mbox{d}yz \right)
\end{displaymath}
which is the expression for the energy of a three dimensional model
that we reported in Ref. \citep{Torrealdea}.

It can be easily checked that the energy derivative $ \dot H =
\nabla H ^T \mathbf{f}_d (\mathbf{x}) $, that is,

 \begin{equation}
\dot{H}=\frac{2 \, p}{\mbox{a}}
\begin{pmatrix}
  \mbox{f}x^2+\frac{\mbox{msd}-\mbox{gnr}}{\mbox{a}}x+\mbox{g}w \\
  \mbox{a}y-\mbox{d}z \\
  \frac{\mbox{d}}{\mbox{ams}}(\mbox{msd}-\mbox{gnr})z-\mbox{d}y\\
  \mbox{g}x
\end{pmatrix}
\begin{pmatrix}
  \mbox{b}x^2-\mbox{c}x^3+\xi \mbox{I}  \\
  \mbox{e}-y  \\
  \mbox{msh}-\mbox{m}z\\
  \mbox{nrl}-\mbox{nk}w
\end{pmatrix},
\end{equation}
is also dimensionally consistent with a dissipation of energy. As the states of an isolated Hindmarsh-Rose neuron are confined to an attractive manifold the range of possible values of its energy is recurrent and the long term average of its energy derivative is zero. However, it has to be considered that the average involves a global balance of energy. The model itself incorporates, in a non explicit way, components which are responsible of the energy consumption together with others which are the energy suppliers.\\

\begin{figure}
\begin{center}
\includegraphics[width=0.60\textwidth]{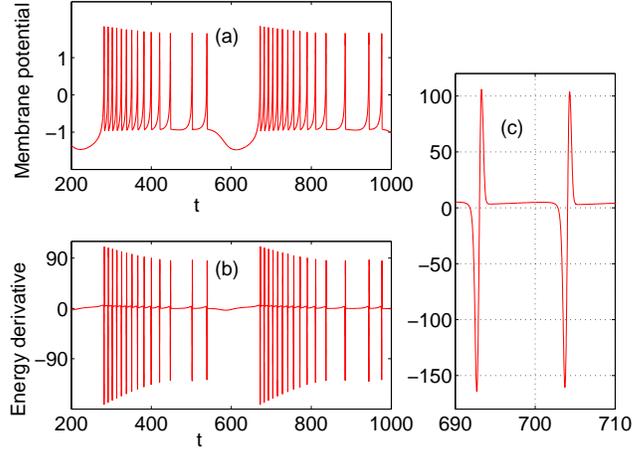}\\
\end{center}
\caption{(a) Action potentials and (b) energy derivative for the
Hindmarsh-Rose model neuron. (c) Detail of the energy derivative
associated to two spikes showing the dissipation of energy during
the depolarization of the membrane potential (negative area) and
its ulterior demand of energy during its repolarization period
(positive area).} \label{fig2}
\end{figure}

Figure 2(a) shows a series of action potentials (variable $x$ in the
model neuron) and Fig. 2(b) shows the energy derivative $\dot H $
corresponding to that series. In Fig. 2(c) a detail of the energy
derivative corresponding to a train of two action potentials is also
shown. For each action potential it can be appreciated that the
energy derivative is first negative, dissipation of energy while the
membrane potential depolarizes during the rising period of  the
spike, and then positive, contribution of energy to repolarize the
membrane potential during its descending period. During the
refractory period between the two spikes the energy derivative
remains slightly positive, still demanding energy, until the onset
of the following action potential. To link the demand of energy to
the repolarizing process is in agreement with the results about the
energetics of neural activity in rat brain by Attwell and Laughlin
\citep{Attwell} who found that, for spiking frequencies of $4$ Hz,
$15\%$ of the energy is used to maintain resting membrane potentials
in non firing epochs while the remaining $85\%$ is used to
restabilize membrane potentials in firing epochs. We calculate the
average energy consumption per unit time of the neuron, that is, the
metabolic energy that has to be supplied to the neuron to maintain
its activity, evaluating the long term average of the negative
component of the energy derivative, that is the energy that is
dissipated in the process of spike generation.

\subsubsection{Two electrically coupled neurons}
Let us consider two Hindmarsh-Rose neurons electrically coupled according to the following equations,
\begin{equation}
\begin{array}{ll}
\dot{x}_i = \mbox{a} \, y_i+\mbox{b} \, x_i^2-\mbox{c} \, x_i^3-\mbox{d} \, z_i+\xi I_i+k_i(x_j-x_i), \\
\dot{y}_i = \mbox{e} -\mbox{f} \, x_i^2-y_i-\mbox{g} \, w_i, \\
\dot{z}_i = \mbox{m}(-z_i+\mbox{s} (x_i+\mbox{h} )), \\
\dot{w}_i = \mbox{n} (-\mbox{k} \, w_i+\mbox{r} (y_i+\mbox{l} )),
\end{array}
\end{equation}
where $k_i \geq 0$ is the coupling strength and $i,j=1,2$\,; $i\neq
j$ are the indices for the neurons. Note that the coupling affects
only to their respective first variables $x_1$ and $x_2$. This kind
of coupling between neurons has been very often reported
\citep{Pinto, Abarbanel,Huerta,Rosenblum,Belykh, Hansel, Dhamala}.

Considering the energy of a neuron given by Eq. (13) and also Eqs. (4) and (5) we have for the average energy variation through the membrane of neuron of index $i$,

\begin{equation}
\langle\dot{H}\rangle_m^i=\frac{2p}{\mbox{a}}
\begin{pmatrix}
  \mbox{f} \, x_i^2+\frac{\mbox{msd}-\mbox{gnr}}{\mbox{a}} \, x_i+\mbox{g} \, w_i \\
  \mbox{a} \, y_i-\mbox{d} \, z_i \\
  \frac{\mbox{d}}{\mbox{ams}}(\mbox{msd-gnr}) \, z_i-\mbox{d} \, y_i\\
  \mbox{g} \, x_i
\end{pmatrix}
\begin{pmatrix}
  \mbox{b} \, x_i^2-\mbox{c} \, x_i^3+\xi I_i  \\
  \mbox{e}-y_i  \\
  \mbox{msh}-\mbox{m} \, z_i\\
  \mbox{nrl}-\mbox{nk} \, w_i
\end{pmatrix},
\end{equation}
where $i,j=1,2$\,; $i\neq j$. As it has been said, the energy consumption of neuron $i$ corresponds to the average of the negative component of this derivative.

The average energy variation at the synapse site of neuron $i$ is given by,

\begin{equation}
\langle\dot{H}\rangle_s^i=\frac{2p}{\mbox{a}} \left (\mbox{f} \, x_i^2+\frac{\mbox{msd-gnr}}{\mbox{a}} \, x_i \right ) \, k_i(x_j-x_i),
\end{equation}
 These equations are used in what follows in the different circumstances in which the  computation of energy  is required.

\section{Information considerations}

A neuron responds to changes in the applied external current and to
inputs from other neurons with changes in its firing regime that
modify its information capacity as well as its average energy
consumption. Shannon's information theory \citep{Shannon} provides a
framework to quantify the amount of information that neurons can
convey during its signaling activity. The first application of
Shannon´s theory to estimate the information entropy of spike
trains was due to MacKay and McCulloch \citep{MacKay}. A
comprehensive approach to understanding the information content of
neural spikes, together with a review of some important
contributions to this area of research can be found in Ref.
\citep{Rieke}.

The information entropy $S$ of a discrete
distribution of probability $p_i$ is defined by $S=-\sum_{i}p_ilog_2
p_i$. This entropy is usually contemplated as a measure of the
average uncertainty of occurrence of events that have a probability
of occurrence $p_i$. Although the information entropy of a discrete
probability distribution is well defined, the situation is less
clear when what is sought is the information entropy of a train of
spikes emitted by a neuron. A long spike train emitted by a neuron can be observed as a succession of windows of spike trains of $T$ ms length which are
partitioned into bins of $\triangle{\tau}$ ms length each. The presence or
absence of a spike inside one of these bins can be codified as $1$ or
$0$ respectively, so that each window of spikes can be contemplated
as a particular symbol from a binary alphabet of $2^{\frac{T}{\triangle{\tau}}}$ different
symbols. We suppose $\triangle{\tau}$ small enough as to count no more than one spike per bin.

Let us suppose that two interneurons are coupled by an electrical synapse according to Eq. (15). Let $T^i_1$ and $T^i_2$ represent symbols of the grammar of the different possible symbols that can be coded with the spike trains of the presynaptic and postsynaptic neurons respectively. The information entropy rate $S_j$ of neuron $j=1,2$ will be,
\begin{equation}
 S_j=-\sum_{i}p(T^i_j)log_2p(T^i_j).
\end{equation}
The mutual information, $I_m$, between the spike trains of both neurons can be defined as
\begin{align}
\nonumber I_m=&-\sum_{i}p(T^i_2)log_2p(T^i_2)\\
             &+\sum_{j}p(T^j_1)\sum_{i}p(T^i_2/T^j_1)log_2p(T^i_2/T^j_1),
\end{align}
that is, the entropy of the postsynaptic train minus the average of its conditioned entropies. This formulation emphasizes the fact that the mutual information between the two trains of spikes can be contemplated as the reduction in entropy that would take place in the postsynaptic train  if the presynaptic one was known.

According to \citep{Nemenman}, since the maximum likelihood estimate of the probabilities $p(T_i)$ is given by the frequencies, Eq. 18 provides an estimate of maximum likelihood of the real entropy. This estimator is biased and underestimates the entropy. With good sampling, $ N \gg K$ with $K$ possible neuron responses and $N$ number of samples, the estimate deviates from the correct value in an additive error ($K-1)/2N$ plus a term proportional to $ 1/N$.

It has to be noted that the mean firing rate of a train of spikes conditions the probabilities of occurrence of the different symbols of the alphabet making their distribution not uniform on the set of bins. This fact reduces the variability of the signal and the actual value of entropy do not reach its maximum possible limit of $T/\triangle{\tau}$ bits per window. The maximum entropy rate that can be measured with time precision $\triangle{\tau}$ from a spike train of mean firing rate $\bar{r}$ is given by \citep{Rieke},
\begin{equation}
S_{max}=\frac{-\bar{r}\triangle{\tau}\,log_2\,(\bar{r}\triangle{\tau})-(1-\bar{r}\triangle{\tau})\,log_2\,(1-\bar{r}\triangle{\tau})}{\triangle{\tau}}
\end{equation}
That is, the firing rate imposes a limit to the maximum entropy rate of a given signal. At a given mean firing rate the maximum entropy is reached when the presence or absence of spikes in a time bin $\triangle{\tau}$ is independent of all other time bins, that is, when there are no temporal correlations in the timing of the spikes. If in addition to the absence of temporal correlations the spike train is perfectly reproducible across repetition of the same stimulus, that is, if there is no noise, this maximum entropy rate sets an upper limit on the information that can be transmitted at the observed spike rate which is termed coding capacity. The actual information transmitted by the spike train compared with its coding capacity provides a measure of the efficiency of the coding.

In this work we have performed a naive estimate \citep{Strong} of the information entropy, generating successive windows of spikes of $25$ ms length which are
partitioned into 5 bins of $5$ ms length each. The presence or
absence of a spike inside one of these bins is codified as $1$ or
$0$ respectively, so that each window of spikes represents a symbol from a binary alphabet of $K=32$ different symbols. To estimate entropies in coupled neurons we have generated $N=10000$ samples. According to Eq. 20 the maximum entropy rate that we could measure is $S=5$ bits per average $25$ ms window in case we had a long spike train firing at 100 Hz mean rate and with no time correlations. In practice, for the cases of coupled neurons we have studied, the active leading neuron is activated with an external current $I=3.024$ what makes the neuron fire at a mean firing rate $\bar{r}=39$ Hz. At this firing rate, according to Eq. 20, the corresponding maximum entropy is $S_{max}=3.65$. Using  Eq. 18, we have estimated $S=3.15$ with an additive error $(K-1)/2N=0.007$. Thus, for the studied cases, windows of spike trains of $T=25$ ms partitioned into bins of $\triangle \tau=5$ ms provide enough variability for the observed neuron responses as to obtain, with no significant error, an information entropy near to its maximum value.

Extrapolation of the windows to larger word length only imply very small corrections that have no incidence in our conclusions. On the other hand, the value of the information entropy rate that is obtained when calculating the entropy of a given spike train is very dependent on the size of the time bin $\triangle \tau$ used for its calculation. The entropy increases as $\triangle \tau$ decreases illustrating the increasing capacity of the train to convey information by making use of the spike timing \citep{Rieke}. The value $\triangle \tau=5$ ms used in this work corresponds to timing each spike to within 20 \% percent of the typical interspike interval of the leading neuron which fires at a mean rate $\bar{r}=39$ Hz. This value of time resolution and the window length used in this work are frequently used with empirical and simulated data. They are computationally appropriate and are used in Ref. \citep{Koch} to explore retinal ganglion cells. Also in Ref. \citep{Strong} similar values are used to analyze responses of a motion sensitive neuron in the fly visual system. For our purpose changing the time resolution $\triangle \tau$ supposes a scale change in the calculated amount of information transmitted between two coupled neurons at different values of their coupling gain and therefore has no effect on the form of the curves ratio of information to energy.

\section{Computational results}
\begin{figure}
\begin{center}
\includegraphics[width=0.6\textwidth]{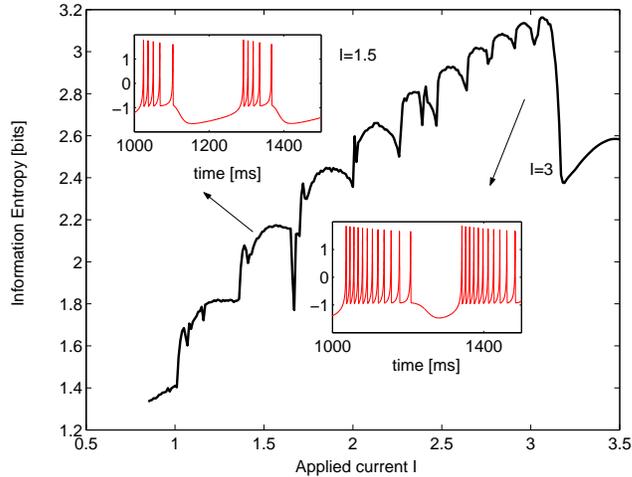}\\
\end{center}
\caption{Information entropy per $25$ ms spike train of an isolated neuron at different values of the external applied current $I$. Computation has been performed generating $2000$ spike trains of $25$ ms at each value of the external current $I$. In the insets bursting regimes corresponding to $I=1.5 \, \textrm{nA}$ and $I=3 \, \textrm{nA}$.} \label{fig3}
\end{figure}
In this section we present results firstly for the neuron
considered as an isolated oscillator and secondly for two neurons
coupled electrically. For the isolated neuron results of energy
consumption and information entropy at different values of the
applied external current $I$ are analyzed. For neurons coupled
electrically, unidirectional and bidirectional couplings have been
studied for identical and nonidentical neurons and results
relative to mutual information, energy consumption, information to
energy ratio and synapse contribution of energy are given.

\subsection{Information entropy and energy consumption in the isolated neuron}
To study the relationship between information entropy and energy
consumption in the different possible oscillatory regimes of an
isolated neuron we have computed its information entropy and average
energy consumption at different values of the applied current $I$.
To quantify the information capacity of the neuron in its different
signalling regimes we have used Eq. (18). The computation has been
performed generating 2000 different spike trains of $25\, ms$ length
at every value of the external current. As it can be seen in Fig. 3,
the information entropy increases in plateaus corresponding to
progressively richer signalling activity. This is so because the
bursting regime of the isolated neuron is very sensitive to its
applied external current $I$. Increasing $I$ gives rise to
subsequent bursting regimes of an increasing number of spikes per
burst \citep{Pinto,Hansel}. The two insets to Fig. 3 show two
examples of bursting regimes corresponding to $I=1.5 \, \textrm{nA}$
and $I=3 \, \textrm{nA}$.

Average energy consumption has been computed averaging over
sufficiently large periods of time the negative part of the energy
derivative given by Eq. (14). Results for consumption are displayed
as positive, i.e. we define consumption as the absolute value of the
energy dissipated. Figure 4 shows the results. The energy
consumption of the neuron increases in steps with $I$, being very
sensitive to the different firing regimes. The different plateaus
correspond to the subsequent bursting regimes of increasing number
of spikes per burst. Figure 4 also shows, in dots, the average
number of spikes that the isolated neuron emits per unit length at
different values of the applied current $I$. As it can be
appreciated the energy consumption is more or less proportional to
the average number of spikes per unit time. In the range of values
of $I$ between $2.5\, \textrm{nA}$ and $3\, \textrm{nA}$ the
linearity is remarkable. A linear relation between energy
consumption and frequency of spikes is what should be expected as
energy consumption is basically linked to the generation of action
potentials. This linear relation has been sometimes hypothesized
in theoretical studies of energy efficiency in the signal
transmission by neurons \citep{Laughlin,Levy96,Levy02}. Our results
show that this simple law does not apply exactly to every signalling
regime in the isolated neuron and, as we show later on, it is not
going to be followed when two neurons are coupled.

\begin{figure}
\begin{center}
\includegraphics[width=0.7\textwidth]{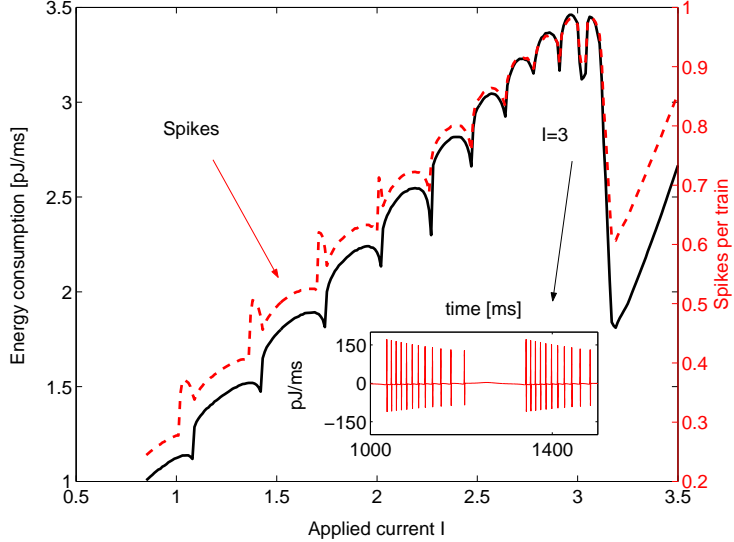}\\
\end{center}
\caption{Energy consumption rate, left vertical axis, and average number of spikes per train, right axis, of an isolated neuron at different values of the external applied current $I$. Computation has been performed generating $2000$ spike trains of $25$ ms at each value of the external current $I$. In the inset temporal energy derivative corresponding to $I=3 \, \textrm{nA}$.} \label{fig4}
\end{figure}

\subsection{Two electrically coupled neurons}

In this section we consider two neurons coupled electrically. We suppose that the presynaptic neuron always signals in a chaotic regime corresponding to a external current $I_1=3.024$. As
we have seen, in this chaotic regime the neuron signals at its
maximum information capacity. In the identical cases the receiving
neuron also signals at $I=3.024$. For the nonidentical cases we
have set the postsynaptic neuron close to its quiescent state at a
low value, $I_2=0.85$, of its external current.  We have analyzed
the unidirectional case setting the gain parameters between both
neurons as $k_1=0$, $k_2=k$ and the bidirectional case setting
$k_1=k_2=k$. Computation has been performed, in every case,
generating 10000 different spike trains of $25 \, ms$ length at
every value of the gain parameter $k$.

\subsubsection{Mutual information}

Using Eq. (19) and the coding explained before, we have computed
the mutual information between the trains of spikes of the pre and
postsynaptic neurons at different values of the coupling strengths
$k_1$ and $k_2$. The mutual information between both neurons as
well as the information entropy rate of the presynaptic and
postsynaptic spike trains are shown in Fig. 5.

When the two neurons are identical, sufficiently large values of
the coupling strength lead both neurons to complete
synchronization and, therefore, to a noiseless channel where no
loss of information takes place. For two identical neurons with
unidirectional coupling results are displayed in Fig. 5(a). At
$k=1$ the two neurons are completely synchronized and their mutual
information reaches its maximum value that corresponds to the
noiseless channel. The constant value of the entropy of the
sending neuron serves as reference. The highest value measured for
the entropy rate of the receiving neuron is $0.18$ bits per second
that corresponds to an entropy of $4.5$ bits per average $25$ ms
train which is very near to the maximum possible entropy value,
$5$ bits per $25$ ms train, that our procedure can detect. Figure
5(b) shows the mutual information rate between two identical
neurons bidirectionally coupled  and the information entropy rate
of the sending and receiving neurons at different values of the
coupling gain $k$. Due to the symmetry of the coupling the
information entropy of both neurons is identical. At $k=0.5$ the
two neurons are completely synchronized and the mutual information
reaches its maximum value that corresponds to the noiseless
channel.

\begin{figure}
\begin{center}
\includegraphics[width=0.7\textwidth]{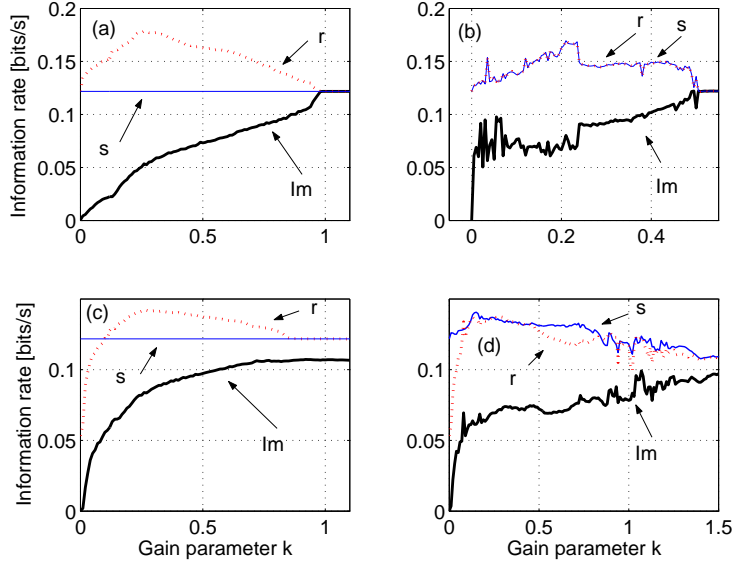}\\
\end{center}
\caption{Mutual information, $Im$, and information entropies of
the sending, $s$, and receiving, $r$, neurons at different values
of the coupling strength $k$. (a) Identical neurons and
unidirectional coupling.  (b) Identical neurons and bidirectional
coupling. (c)Nonidentical neurons and unidirectional coupling. (d)
Different neurons and bidirectional coupling.} \label{fig5}
\end{figure}

In practice actual channels are always noisy and neurons
nonidentical and it is of practical interest to know about the
efficiency of the signal transmission in these circumstances. For
nonidentical neurons with unidirectional coupling results are
shown in Figure 5(c). As it can be appreciated, the information
entropy of the receiving neuron increases rapidly with $k$
reflecting the fact that the signalling activity of the
postsynaptic neuron becomes more complex due to the coupling.
Eventually, at larger values of $k$, the information entropy of
the receiving neuron equals the entropy of the sender in spite of
the fact that both neuron are nonequal and complete synchrony does
not take place. The mutual information also starts a slower and
progressive increase from being zero when both neurons work
independently, at $k=0$, to its maximum value slightly larger than
$0.1 \, bits/ms$ which is reached at about $k=0.7$. As the channel
is noisy the mutual information never reaches the information
entropy of the presynaptic signal. Finally, Fig. 5(d) shows the
mutual information rate and the entropy rates of two nonidentical
neurons coupled bidirectionally. The dynamics of the neurons is
now more complex. The information entropy of the receiving neuron
soon reaches a relatively high value and progressively diminishes
with subsequent increment of the coupling strength $k$. The
information entropy of the sender also experiences a slight
decline as both neurons mutually synchronize. The channel is
noisier than in the unidirectional case and the values reached by
the mutual information are now lower and more erratic.

\subsubsection{Average energy consumption}
We have calculated the average energy consumption per unit time
that is required to maintain the organized bursting of two
electrically coupled neurons. To produce these results the
negative component of the energy variation given by Eq. (16) has
been averaged over $10000$ different spike trains of $25 \, ms$
length. Figure 6 shows the results at different values of the
coupling strength $k$.

Results for two identical neurons when the coupling is
unidirectional are shown in Fig. 6(a). It becomes apparent that
there is a region of values of the coupling parameter, around
$k=0.6$ where signalling occurs at minimum values of energy
consumption by the receiving neuron. The energy consumption of the
sending neuron remains constant because its dynamics is not
affected by the unidirectional coupling. In Fig. 6(b) we can see
what happens to the same identical neurons when the coupling is
bidirectional. Due to the symmetry of the coupling the energy
average consumption of the sending and receiving neurons are
practically the same. It is remarkable the very neat reduction of
energy consumption that takes place between values of the gain
parameter $k$ in the interval $0.2<k<0.25$. Subsequent increases
in the value of the gain maintain the consumption in a flat
plateau at high values of energy consumed. At values of the gain
parameter close to $k=0.5$ the consumption of energy falls to its
initial uncoupled level as complete synchronization takes place.

\begin{figure}
\begin{center}
\includegraphics[width=0.7\textwidth]{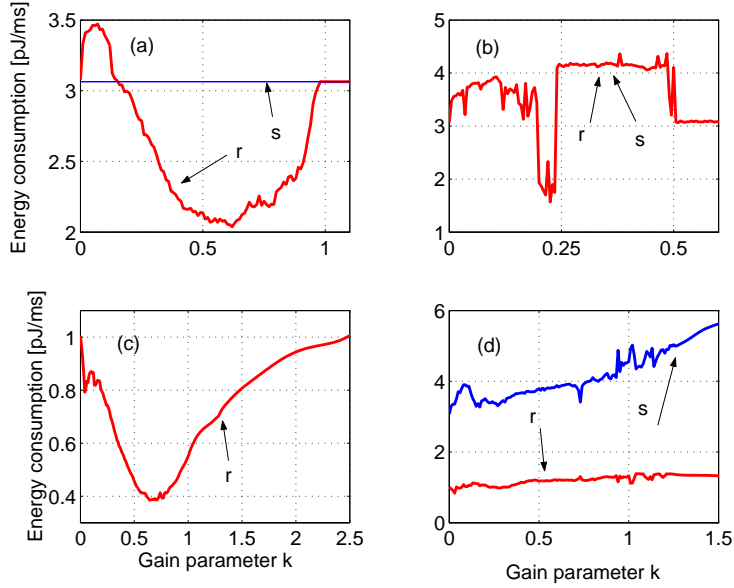}\\
\end{center}
\caption{Average energy consumption through the membrane of the
sending $s$ and receiving $r$ neurons at different values of the
coupling strength $k$. (a) Identical neurons and unidirectional
coupling.  (b) Identical neurons and bidirectional coupling, both
consumptions are equal. (c)Nonidentical neurons and unidirectional
coupling. (d)Nonidentical neurons and bidirectional coupling.}
\label{fig6}
\end{figure}

For nonidentical neurons Fig. 6(c) displays the energy consumption
of the receiving neuron when the coupling is unidirectional. The
constant energy consumption of the sending neuron is not displayed
for scale reasons. The energy consumption of the receiving neuron also exhibits a clear minimum that occurs for a value of the coupling strength around $k=0.6$. Subsequent increase in he gain
$k$ leads to higher and higher levels of energy consumption.
Finally, in Fig. 6(d) we can see the same two nonidentical neurons
when the coupling is bidirectional. As it can be appreciated, the
signaling of the  sending neuron takes place at higher values of
average energy consumption than the ones of the receiving neuron.
This result is quantitatively consistent with the data of average
consumption of energy at different values of the external current
$I$ that we have presented in Fig 4. Despite the unidirectional
case, where the average consumption of energy of the sending
neuron remains constant, in the bidirectional case the coupling
also affects the sending neuron and makes it modify its energy
consumption as a function of the coupling strength $ k$. The
average energy consumption of both neurons follows quite an
irregular pattern which can not be clearly appreciated due to the
scale of the figure.

\begin{figure}
\begin{center}
\includegraphics[width=0.7\textwidth]{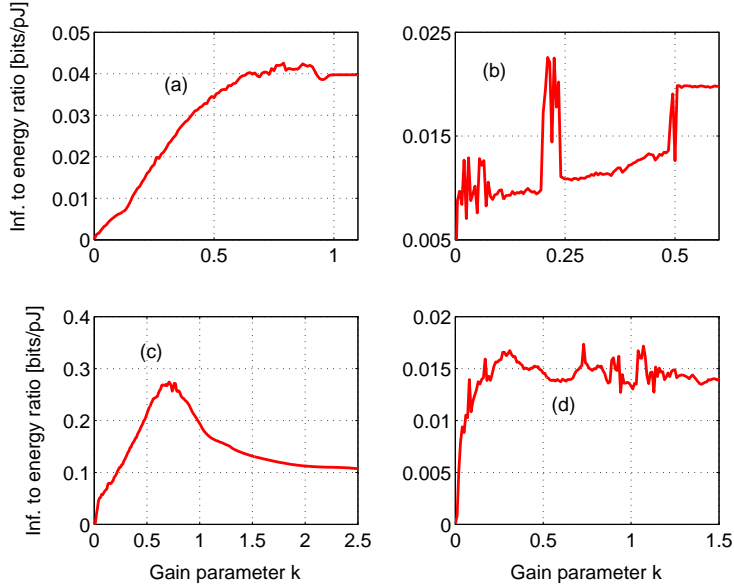}\\
\end{center}
\caption{Mutual information to energy consumption ratio at
different values of the coupling strength $k$. In the
unidirectional coupling the computed energy consumption refers to
the energy consumed through the membrane exclusively by the
receiving neuron. In the bidirectional cases the consumption of
energy refers to the summation of the energy consumed through the
membrane by each of the neurons (a) Identical neurons and
unidirectional coupling.  (b) Identical neurons and bidirectional
coupling. (c) Nonidentical neurons and unidirectional coupling.
(d) Nonidentical neurons and bidirectional coupling.} \label{fig7}
\end{figure}

\subsubsection{Mutual information to average energy consumption ratio}

Figure 7 shows the ratio of mutual information to average energy
consumption rate at different values of the coupling gain $k$, and
Fig. 8 their corresponding averaged synchronization errors measured
as the norm of the error vector
$\mathbf{e}=\|\mathbf{x}_2-\mathbf{x}_1\|$, for the same neurons and
couplings presented before. In Fig. 7(a) the neurons are identical
and the coupling unidirectional. As it can be appreciated, from
$k=0.6$ the information to consumption ratio reaches values even
higher than the one that corresponds to identical synchronization at
$k=1.0$.  As it can be seen in Fig. 8, identical unidirectional
coupling neurons at values of the coupling around $k=0.6$  have
still  a high synchronization error and it is remarkable that
transmitting information at values of the coupling where the channel
remains noisy is energetically more efficient than transmitting with
the noiseless channel that would correspond to complete
synchronization at $k=1$. Figure 7(b) shows the same identical
neurons coupled bidirectionally. In this case both neurons respond to changes in the coupling $k$ with changes in their energy consumption, see Fig. 6(b), and the ratio has been computed
adding the consumptions of the sending and receiving neurons. As it
can be appreciated, between $k=0.2$ and $k=0.25$ the information to
consumption ratio has a neat peak with very high relative values. In
this region of the coupling strength, the information to consumption
ratio reaches values even higher than the one that corresponds to
identical synchronization at $k=0.52$. Noticeably enough, at those
values of the coupling strength the synchronization error is
maximum, Fig. 8.  This is due to the fact that in that region of
values of the coupling strength both neurons synchronize in
anti-phase  producing large synchronization errors and maximum
correlation values \citep{Torrealdea}.
\begin{figure}
\begin{center}
\includegraphics[width=0.7\textwidth]{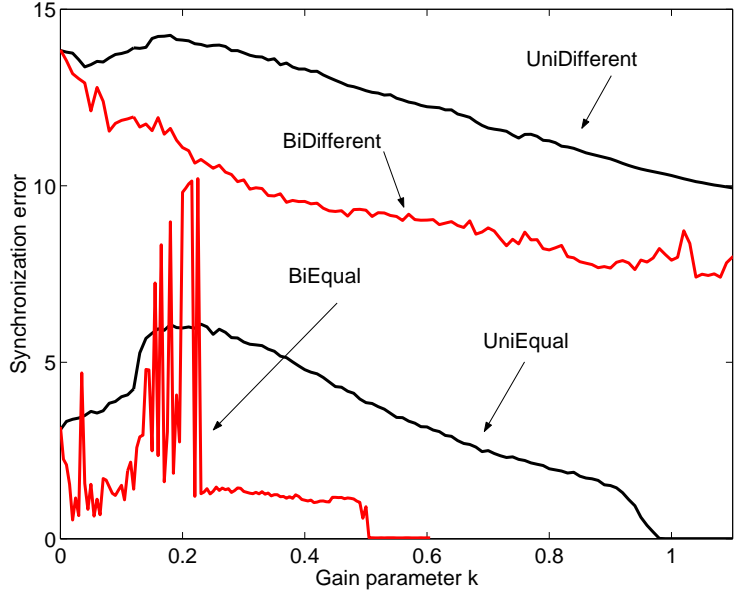}\\
\end{center}
\caption{Average synchronization error between two coupled neurons
at different values of the coupling gain $k$. The synchronization
error has been measured as the norm of the error vector
$\mathbf{e}=\|\mathbf{x}_2-\mathbf{x}_1\|$ and the average has
been done, for each value of $k$, over $10000$ trains of $25 \,
ms$ length. The curves labelled \emph{UniEqual} and \emph{BiEqual}
show results for the case of identical neurons with coupling
respectively unidirectional and bidirectional and the curves
labelled \emph{UniDifferent} and \emph{BiDifferent} show the
respective results for nonidentical neurons. The external applied
current is always $I=3.024$ for the sending neuron and $I=3.024$
or $I=0.85$ for the receiving neuron in the identical and
nonidentical cases respectively.} \label{fig8}
\end{figure}
Moreover, anti-phase synchronization occur at the least possible
values of energy consumption, Fig. 6(b), what altogether produces
the highest efficiency of the transmission from the point of view
of its energy cost.

Figure 7(c) shows the mutual information rate to average energy
consumption ratio between two nonidentical coupled neurons with
unidirectional coupling. As we have seen before, there is  a
region of values of the coupling strength, at about $k=0.8$, where
the consumption of energy is minimum, Fig. 6(c), and the mutual
information has already reached its maximum possible value, Fig.
5(c). This fact is reflected in Fig. 7(c) where the maximum value
of the mutual information to consumption ratio occurs at $k=0.8$.
Further increase in the coupling strength does not lead to any
improvement in the information transmission rate but to a loss in
the energy efficiency of the transmission. As it can be
appreciated in Fig. 8, the synchronization error for this case at
$k=0.8$ remains high but has no particular influence in the
information to consumption ratio. Figure 7(d) presents the
bidirectional nonidentical case. In this bidirectional case, the
energy consumption corresponds to the total average energy
consumed by both neurons. The ratio of the mutual information to
the total energy consumption of both neurons soon reaches a kind
of uneven plateau with many peaks. All these peaks represent
relative maxima of the information to energy ratio which provide
plenty of room for energy efficient information transmission.

\subsubsection{Relative weight of the contribution from the synapse to the average income of energy through the
membrane}

When a neuron is signalling in isolation the average energy
dissipated through the membrane, energy consumption, is perfectly
balanced by its average income of energy, that is the long term
temporal average of Eq. (14) is zero. Nevertheless, when the
neuron receives a synaptic junction its oscillatory regime is
altered influenced by the synapse. The new oscillatory regime
requires the synapse to play a role in the energy balance that
makes the new regime possible.

\begin{figure}
\begin{center}
\includegraphics[width=0.6\textwidth]{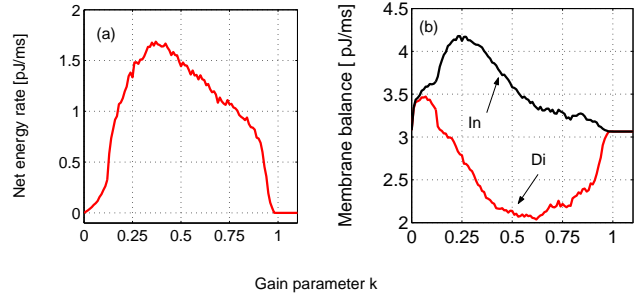}\\
\end{center}
\caption{(a) For two identical coupled neurons net average energy
rate of the receiving neuron at different values of the coupling
gain $k$. The coupling is electrical and unidirectional. (b)
Balance of energy in the receiving neuron. Curve \textit{Di},
average of the negative part of its energy derivative, that is,
the energy consumption rate of the neuron, energy that is
dissipated through its membrane. Curve \textit{In} average of the
positive part of the energy derivative, that is, the total income
of  energy  trough the membrane of the cell. Computation has been
performed generating $10000$ spike trains of $25$ ms at each value
of the coupling gain $k$} \label{fig9}
\end{figure}

Figure 9 (a) shows for two identical unidirectionally coupled
neurons the average energy derivative, that is the long term
average of Eq. (14), of the postsynaptic neuron at different
values of the coupling gain $k$. At $k=0$, when the neuron is
still uncoupled, the average energy derivative is zero, reflecting
the fact that the neuron is signaling according to its natural
isolated dynamics. In that conditions, the average energy received
through the membrane is perfectly balanced by its average
dissipation of energy. As soon as the coupling is working the
dynamics of the neuron changes and this balance is broken. As
shown in Fig. 9(a) the average energy derivative at each value of
$k$ is positive indicating that there is a net and sustained
increase in time of the energy of the neuron or, in other words,
that the average income of energy is larger than the average
energy dissipated trough the membrane. To make more visible the
unbalance of the flow of energy through the membrane that takes
place when a gap junction is present Fig. 9 (b) shows both the
average energy rate dissipated trough the membrane by the
postsynaptic neuron, that is its consumption of energy, and its
average income of energy trough the membrane. As it can be
appreciated, the average power dissipated through the membrane
only equals the average power received trough it when both neurons
are isolated at $k=0$ or when both neurons are completely
synchronized at $k\geq 1$. When synchronization is not
complete the average power supplied through the membrane is always
larger than the average power dissipated through it.

However, the dynamics  of the coupled neuron returns again and
again to the same recurrent regions of the phase space, see Fig.
1, what means that its average energy does not increase in time.
This energetically sustained signalling regime is possible because
the average increase of energy of the receiving neuron is
compensated by a net outflow of energy through the synapse.
Accordingly with Section 2.2 of this paper, our interpretation is
that this energy is somehow transformed and contributed again as
part of the total income of energy trough the membrane, i.e., the
synapse must be energetically active during the synchronized
behavior. Figure 10 shows for every studied case, at different
values of the coupling strength $k$, the relative weight of the
contribution of the synapse to the total energy income of the
receiving neuron through its membrane. For identical
unidirectional neurons, Fig. 10(a), the contribution from the
synapse increases smoothly reaching its maximum contribution,
forty percent of the total at $k=0.4$, and then smoothly
decreasing towards zero as the neurons completely synchronize. In
the bidirectional case, Fig. 10(b), there is no substantial
contribution from the synapse except in the region of values of
the gain parameter $0.2\leq k\leq 0.25$ where the maximum
information to energy ratio takes place. In this region the
contribution from the synapse is as high as sixty percent of the
total income rate of energy to the neuron through its membrane.
For the nonidentical case, Figs. 10(c) and 10(d), the contribution
of the synapse reaches a plateau and becomes independent of the
gain $k$. In both cases the synapse contributes nearly ninety
percent of the total income of energy through the membrane.

These results show that some production of energy at the synaptic
site seems to be necessary for the neuron to keep its coordinated
signalling regime. Nevertheless, there is biological evidence that
links the generation of metabolic energy to the inflow of glucose
through the membrane to produce ATP. Both facts could be reconcile
assuming that the electrical energy produced at the synaptic site
is conveniently transformed and reabsorbed by the neuron through
its membrane for the generation of new spikes. Our proposed global
flow of energy has been schematized in Fig. 11.
\begin{figure}
\begin{center}
\includegraphics[width=0.7\textwidth]{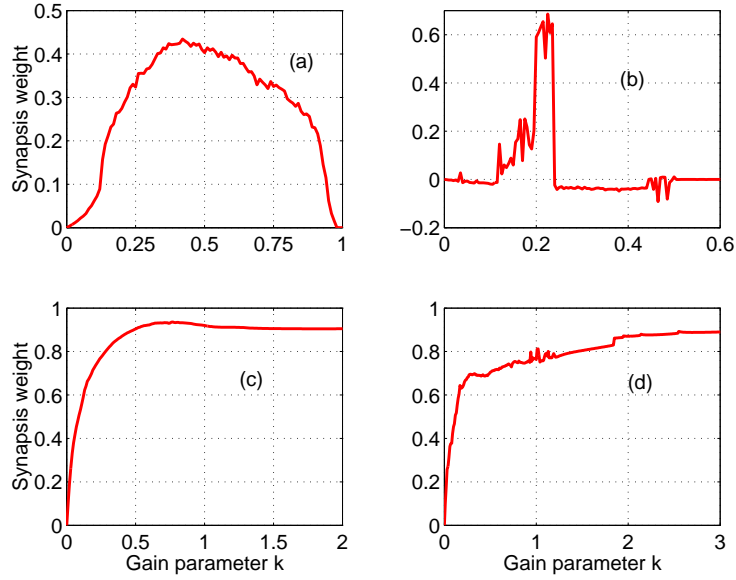}\\
\end{center}
\caption{Relative weight of the contribution of the synapse to the
total income of energy through the membrane of the receiving
neuron at different values of the coupling strength $k$. (a)
Identical neurons and unidirectional coupling. (b) Identical
neurons and bidirectional coupling. (c) Nonidentical neurons and
unidirectional coupling. (d) Nonidentical neurons and
bidirectional coupling.} \label{fig10}
\end{figure}

\section{Discussion and conclusions}
Since the work of Hodgkin and Huxley \citep{Hodgkin}, who where able
to describe the membrane currents of the squid axon via a
quantitative model in differential equations, models of that type
are frequently used
\citep{Rulkov,DeMonte,Ivanchenko,Venaille,Abarbanel,Huerta,Rosenblum,Belykh,
Hayashi,Lago,Yoshioka, Hasegawa,Nagai} to generate and analyze spike
trains with temporal characteristics similar to the ones emitted by
real neurons. We have shown in Refs. \citep{Torrealdea,
Torrealdea07}, that this type of models can also tell us about the
energy implications of producing spike trains. In this work we have
assigned an energy function $H(\mathbf{x})$ to a four dimensional
Hindmarsh-Rose neuron. This function has the characteristics of a
real physical energy and, therefore, it can be used to estimate the
energetic cost of any particular signalling regime, providing the basis
for all the computations involving energy provision or consumption.
We do not imply that this energy function is quantitatively and
accurately describing the changes in energy associated to the
dynamics of a real neuron. What we imply is that if a particular
kinetic model is considered able to describe some dynamical aspects
of the signaling patterns of real neurons and we can
associate to it a function that satisfies some required
conditions, this function represents a physical energy for the model able
to describe the energy implications of its dynamics and, consequently, able to describe some energy implications of the signalling patterns of real neurons. Our approach is valid for many of the frequently used models of neurons in continuous differential equations. In principle the approach is not applicable to models of the type integrate and fire as they do not provide any structural hypothesis to support the election of an appropriate energy function. In Ref. \citep{Ozden} the synchronization between an electronic oscillatory circuit and a real neuron from the inferior olivary nucleus of the rat brain has been reported. To accommodate the oscillation between the circuit and the neuron an electronic coupling consisting of adjustable gain amplifiers is used. Experiments of this type seem to support that a flow of energy must be provided by the coupling mechanism and could be used to obtain information of the energy required for the synchronization of real neurons.

\begin{figure}
\begin{center}
\includegraphics[width=0.7\textwidth]{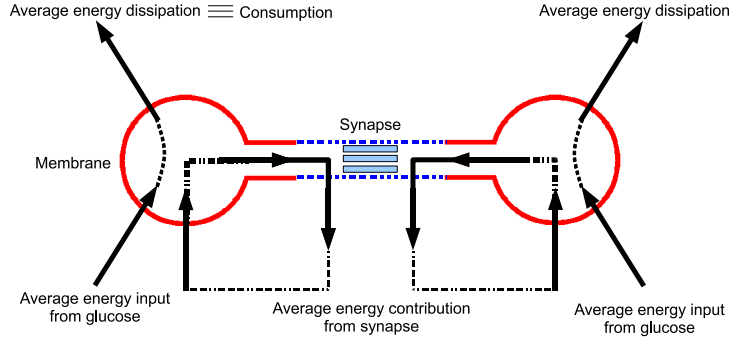}\\
\end{center}
\caption{Average flow of energy through the membranes and at the
synaptic sites in two electrically coupled neurons.} \label{fig11}
\end{figure}

A natural way to propagate information can be through a channel of
electrical coupled neurons were signals are transferred from one
neuron to another. For instance, electrical synapses between AII
amacrine cells and ON-cone bipolar cells are considered essential
for the flow of visual signals in the retina under dark-adapted
conditions \citep{Kolb}. In the transmission of information
synchronization seems to play a key role. Veruky and Hartveit in
Ref. \citep{Veruki2002b} show that spike generation between pairs of
AII amacrine cells can be synchronized precisely and that there is
evidence that spikes can be transmitted through electrical synapses
between these cells. Also in Ref. \citep{Veruki2002a} they
demonstrate temporally precise synchronization of subthreshold
membrane potential fluctuations between amacrine cells and ON-cone
bipolar cells. Identical neurons can always reach precise
synchronization at sufficiently large values of the gain parameter.
Thus, a channel of identical neurons at large enough values of the
synaptic coupling constitutes a noiseless channel where mutual
information reaches its maximum rate and maximum efficiency in the
transmission could be expected. Nevertheless, our results show that
the information to consumption ratio reaches high values, even
higher than the one that corresponds to identical synchronization,
for values of the coupling strength below the ones producing
identical synchronization. Transmitting at these conditions is
energetically advantageous without implying a significant loss in
the information rate. At these values of the coupling strength the
synchronization error is still high and it is remarkable that
transmitting information at values of the coupling where the channel
remains noisy is energetically more efficient than transmitting with
a noiseless channel. If the neurons are nonidentical synchronization
is never complete and the channel is always noisy independently of
the value of the coupling strength. Our results show that there is a
region of values of the coupling strength where the mutual
information is high and the consumption of energy is still
relatively low. Further increase in the coupling strength does not
lead to any improvement in the information transmission rate but to
a loss in the energy efficiency of the transmission.

The existence of regions of high mutual information rate with relatively low consumption of energy can be linked to the appearance of temporal antiphase synchronization. In these regions cross correlation of
instantaneous values shows that the consumption of energy of both neurons is basically incoherent \cite{Torrealdea} what could facilitate a cooperative behavior, especially in a large net of assembled neurons, and questions the point raised by \citep{Lennie} relative to the severe limitations that the high cost of a single spike imposes on the number of neurons that can be substantially active concurrently.

For the central nervous system it has been proposed the existence of
a specialized structural site, for glycolytic generation of ATP,
localized at the postsynaptic site \citep{Wu,Siekevitz}. According
to Ref. \citep{Kasischke} the temporal pattern of the presumed
glycolytic response would directly follow the presynaptic input in
order to meet metabolic needs induced by the processing of nerve
signal transduction. We have found that the synapse must be
energetically active during the synchronized behavior. On average,
there is no flow of energy through the synapse from one neuron to
the other but a flow of energy leaving the neuron at the synaptic
site. It can certainly be instantaneous flows of energy from one
neuron to the other, in fact, it is  believed that the electrical
coupling itself is caused by flow of current through gap junctions
\citep{Veruki2002a}, but the net average flow of energy between the two neurons is zero. The average income of energy through the membrane exactly matches the average output of energy through the
membrane, energy consumption, plus the average energy leaving the
neuron at the synaptic site. We hypothesize that the energy leaving
the neuron at the synaptic site to the extracellular medium does not
substantially degrade and it is somehow fed back again into the
neuron through its membrane, Fig. 11. The Hindmarsh-Rose model of
the dynamics of the neuron does not provide enough biological
information as to be able to decide which terms in the energy
derivative should be considered energetically conservatives. We have
assumed that the net energy contributed from the synapse does not
imply a net energy consumption and that it is in some way recovered
an indefinitely reused for the generation of new spikes. It could be
well the case that part of the synapse energy also degraded. In that
case, the consumption of energy of the neuron to maintain its
signalling activity would have to include the dissipation of energy
in the synapse.

Energy efficient information transmission from the point of view that inputs are optimally encoded into Boltzmann distributed output signals has been analyzed in \citep{Balasubramanian}. An open question is the determination of the energy cost of generating the spike trains that codify each of the different output symbols. Our approach provides a way to determine the energy cost of the generation of different spike trains. It is to be emphasized that the distribution of energy cost of a set of symbols can be very dependent on the particular coupling conditions of the signalling neuron.

When the availability of energy is a significant constraint a
trade-off between  the transfer rate of information between neurons and its energetic cost is to be expected in order to obtain an efficient use of energy by the neurons. Our results, obtained from a comprehensive single model of
neuron that links information and energy, provide room for such a
kind of trade off and suppose a novel approach to the open problem
of whether biological computation optimizes the use of energy in the
generation and transmission of neural codes.
It seems likely that real neurons use energy efficient circuits to
generate and transmit information. It has been reported \citep{Vincent} that the neural organization observed in the early visual system is compatible with an efficient use of metabolic energy. The center surround organization of retinal ganglion cells optimizes the use of  energy when encoding natural images. Other aspects of the organization such as the higher densities of receptive fields in the fovea that decrease in the periphery could also be in an attempt to optimize the use of metabolic energy \citep{Vincent}. In order to test their energy efficient coding hypothesis Vincent et al. use a model where the metabolic cost of the synaptic activity and firing rates can be fully manipulated. In the retinal stage, a cost that increases in proportion to synaptic activity is imposed while in the cortical stage they suppose a cost proportional to the firing rate of neurons. Although this is certainly a plausible hypothesis it is not based on any comprehensive model of energy linked to the true dynamics of the firing regime of the neurons. We think that models of energy like the one described in this paper could provide support to empirical studies to ascertain if neurons really are taking advantage of efficiency savings.

\bibliographystyle{agsm}

\bibliography{HindmarshRose}

\end{document}